\pdfoutput=1
\documentclass[a4paper,11pt]{article}
 \usepackage{lscape}
 \usepackage[all]{xy} 
 \usepackage{longtable} 
 \usepackage{jheppub} 
 \usepackage{graphicx}
 \usepackage{amssymb}
 \usepackage{amsmath}
\usepackage{mathtools}
\usepackage{listings}
\usepackage{dcolumn}
\usepackage{bm}
\usepackage{color}
\usepackage{multirow}
\usepackage{upquote} 
 
\newcommand{\met}{{/\!\!\!\! E_T}} 
 
\allowdisplaybreaks

\usepackage[table]{xcolor}

\definecolor{mag}{RGB}{255,0,255}
\definecolor{cy}{RGB}{0,255,255}
\definecolor{blu}{RGB}{51,51,255}

\newcommand{\lsim}{{\;\raise0.3ex\hbox{$<$\kern-0.75em\raise-1.1ex\hbox{$\sim$}}\;}}
\newcommand{\gsim}{{\;\raise0.3ex\hbox{$>$\kern-0.75em\raise-1.1ex\hbox{$\sim$}}\;}}
\newcommand{\beq}{\begin{equation}}
\newcommand{\eeq}{\end{equation}}
\newcommand{\bea}{\begin{eqnarray}}
\newcommand{\eea}{\end{eqnarray}}
\newcommand{\DF}{\Delta_{4}}
\mathchardef\minus="002D

\title{\boldmath Enhancing the discovery prospects for SUSY-like decays with a forgotten kinematic variable}
\author[a,b]{Dipsikha Debnath,}
\author[c]{James~S.~Gainer,}
\author[d]{Can Kilic,}
\author[e,f]{Doojin Kim,} 
\author[a]{Konstantin~T.~Matchev,}
\author[d]{and Yuan-Pao Yang}

\affiliation[a]{Physics Department, University of Florida, Gainesville, FL
  32611, USA}
\affiliation[b]{NHETC, Department of Physics and Astronomy, Rutgers, The State University of NJ, Piscataway, NJ 08854, USA}
\affiliation[c]{Department of Physics and Astronomy, University of Hawaii,
  Honolulu, HI 96822, USA}
\affiliation[d]{Theory Group, Department of Physics,
  The University of Texas at Austin, Austin, TX 78712, USA} 
\affiliation[e]{Theoretical Physics Department, CERN, Geneva, Switzerland} 
\affiliation[f]{Department of Physics, University of Arizona, Tucson, AZ 85721, USA}

\abstract{The lack of a new physics signal thus far at the Large Hadron Collider motivates us to consider how to look for challenging final states, with large Standard Model backgrounds and subtle kinematic features, such as cascade decays with compressed spectra. Adopting a benchmark SUSY-like decay topology with a four-body final state proceeding through a sequence of two-body decays via intermediate resonances, we focus our attention on the kinematic variable $\DF$ which previously has been used to parameterize the boundary of the allowed four-body phase space. We highlight the advantages of using $\DF$ as a discovery variable, and present an analysis suggesting that the pairing of $\DF$ with another invariant mass variable leads to a significant improvement over more conventional variable choices and techniques.}

\preprint{
\begin{flushleft} 
UTTG-09-18\\
CERN-TH-2018-203\\
UH-511-2018-1295
\end{flushleft} 
}

\begin{document} 
\maketitle
\flushbottom


\section{Introduction}
\label{sec:introduction}

\begin{figure}
\begin{center}
\includegraphics[width=10cm]{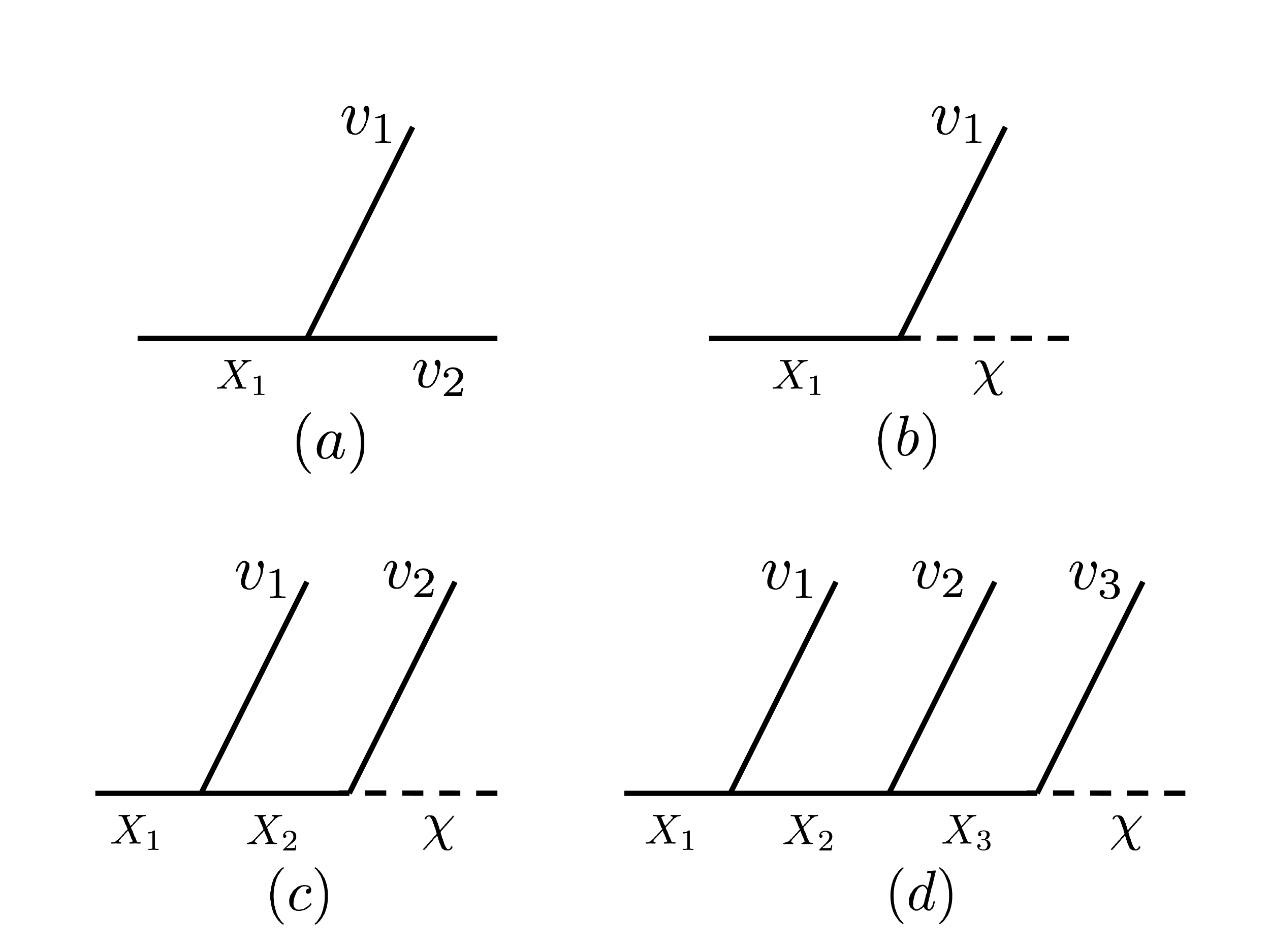}
\end{center}
\caption{Benchmark decay topologies which allow for inclusive searches for the production of a new heavy resonance $X_1$. 
Here $v_1$, $v_2$ and $v_3$ are SM particles which are reconstructed in the detector
(either directly, or through their respective visible decay products), while
$\chi$ is a potential dark matter candidate which is invisible in the detector.
$X_2$ and $X_3$ are additional BSM particles with masses $m_{X_1}>m_{X_2}>m_{X_3}>m_{\chi}$. }
\label{fig:cascade}
\end{figure}

The possible existence of particles beyond the Standard Model (SM) at the TeV scale is theoretically motivated both by naturalness considerations for the electroweak scale~\cite{Gildener:1976ai}, 
and by the so-called WIMP (weakly interacting massive particle) miracle for obtaining the correct dark matter relic abundance~\cite{Jungman:1995df}. 
Nevertheless, as we approach the end of Run II of the Large Hadron Collider (LHC), we have as yet no conclusive evidence of new particles beyond the SM (BSM)~\cite{recent}. 
This requires us to pause, rethink and perhaps re-optimize our search strategies, in preparation for what may lie ahead. In particular, we should 
be mindful of the following challenges:
\begin{itemize}
\item {\em The signal may be buried under a large SM background.} Of course, one obvious possibility for why partner particles may so far have evaded detection 
is that they are simply too heavy and therefore have small production cross sections. If that is the case, then discovery could be waiting around the corner, 
provided that the signatures of the new particles are distinctive. For instance, significant mass gaps in the spectrum of the new particles will result in high $p_T$ leptons and jets 
in the final state and a sizable missing transverse energy, $\met$. Therefore, while the signal cross section may be low, signal over background 
can still be large and reaching discovery sensitivity will simply be a question of collecting sufficient statistics. This scenario is rather uninteresting to us, 
and instead in this paper we focus on the alternative --- that the new particles are being produced in sizable numbers, but their signatures are 
plagued by large SM backgrounds, so the name of the game is whether we can identify selection criteria which have the best potential for discriminating against the background.
This attitude is supported by the flurry of theoretical activity in recent years in designing models which ``hide" the new physics from the LHC.
One of the standard methods for doing so is to arrange for a ``compressed" mass spectrum with a mass degeneracy 
of the relevant particles, such as supersymmetric (SUSY) partners, so that the resulting decay products are too soft to be triggered upon and tagged 
in the experimental analysis \cite{Martin:2007gf,Baer:2007uz,Schwaller:2013baa,Han:2014kaa,Han:2014aea,An:2016nlb,Konar:2016ata,Zarucki:2017imv,Aaboud:2017leg,Schofbeck:2016oeg}, or a ``stealth" mass spectrum, 
where the new physics signature becomes identical to the SM background,
since the additional particles are too soft to make any appreciable difference
\cite{Fan:2011yu,Alves:2012ft,Cho:2014yma,Macaluso:2015wja,Cheng:2016npb,CMS:2014exa,ZeviDellaPorta:2016nan}. 
Our aim will be to highlight a kinematic variable that, either by itself or in conjunction with more conventional variables, 
can more effectively select signal over background when the signal spectrum is compressed and when signal events contain multi-stage cascade decays.

\item {\em Exclusive searches may be reducing the signal statistics to unobservable levels.} When searching for new physics, one has to find the right balance between
inclusive and exclusive searches. Inclusive searches are more robust since they have fewer theoretical assumptions about the event topology and have a higher signal efficiency. 
On the flip side, they tend to suffer from larger SM backgrounds. In contrast, exclusive searches have the potential to reach higher sensitivity when the correct assumptions are made about the features of signal events, since those features can then be used to reduce backgrounds, but at the cost of relying on the assumptions about event topology that may prove to be incorrect.

In our study we will remain much more inclusive than in experimental searches that model the topology of the {\it entire} event, and instead we will only operate on the assumption that the event contains (at least) one SUSY-like cascade decay proceeding through a sequence of two-body decays and with an invisible particle at the end of the decay chain. We will make no assumptions about whether the particle at the beginning of the cascade is singly or pair-produced, and if the latter, what the ``other side'' of the event looks like. Because of this, we will not make direct use of $\met$, or any other transverse variables. Adopting a benchmark final state with three visible and one invisible final state particles [see Fig.~\ref{fig:cascade}(d)], we will focus our attention on fully Lorentz-invariant kinematic variables. 

\item {\em Uncertainties in background modeling.} A required component of any new physics search is the prediction of the
expected SM background. Depending on the final state, this may turn out to be a difficult task, plagued by 
large systematics. Ideally one would like to use data-driven background estimates, and not rely on theoretical input or Monte Carlo.
The classic technique for such searches is the ``bump hunting" method with sideband subtraction. Fig.~\ref{fig:cascade}(a-c) shows examples 
of simpler decay chains for which this method is easily applied. Fig.~\ref{fig:cascade}(a) depicts a visibly decaying resonance, here to 
two visible particles $v_1$ and $v_2$. In this case, the relevant kinematic variable is the invariant mass $m_{v_1v_2}$ of the decay products --- 
it exhibits a Breit-Wigner peak at the mass $m_{X_1}$ of the new resonance. Since the $m_{v_1v_2}$ distribution for the SM background is expected to be smooth, one can 
interpolate from the sidebands and obtain a reliable prediction for the background under the peak. This tried-and-true method has been used successfully many times in the past, 
including most recently for the discovery of the Higgs boson in the diphoton channel~\cite{Aad:2012tfa,Chatrchyan:2012xdj}. 

However, the method runs into a complication if one of the final state particles is invisible in the detector, e.g. particle $\chi$ in Fig.~\ref{fig:cascade}(b). 
Nevertheless, the procedure still goes through, only this time one has to use a suitable kinematic variable which retains the ``bump" feature for the signal,
namely the {\em transverse} invariant mass $m_{T,X_1}$ \cite{Smith:1983aa,Barger:1983wf,Betancur:2017kqe}. The downside of the transverse mass variable $m_T$ (and the related 
mass variables $m_{T2}$~\cite{Lester:1999tx}, $m_2$~\cite{Barr:2011xt,Mahbubani:2012kx,Cho:2014naa}, etc.) is that its definition uses the $\met$ measurement, which forces a departure from inclusivity, and also suffers from the systematics of all possible detector effects. For decay chains containing more than one visible particle, one can remain more inclusive by working only with Lorentz-invariant variables constructed from the momenta of these particles. For the two-stage decay chain in Fig.~\ref{fig:cascade}(c), the only such kinematic variable is the invariant mass $m_{v_1v_2}$, whose distribution does have a distinctive feature \cite{Cho:2015nxy}. While these cases have all been studied in great detail in the past, there has not been a comparable effort to design optimized variables for a longer decay chain, such as in Fig.~\ref{fig:cascade}(d). We will therefore adopt this decay topology as our benchmark in this paper. Our main goal will be to identify and study a kinematic variable for this decay topology that is robust to a certain amount of uncertainty in the modeling of the relevant backgrounds.
\end{itemize}

Based on the arguments above, an obvious choice of kinematic variables to consider are the 
pair-wise\footnote{The invariant mass variable $m_{v_1v_2v_3}$ of all three visible particles is not an independent quantity, since
$$m_{v_1v_2v_3}^2=m_{v_1v_2}^2 + m_{v_2v_3}^2 +m_{v_1v_3}^2 -m_{v_1}^2  -m_{v_2}^2  -m_{v_3}^2.$$ } 
invariant masses of the visible decay products, $m_{v_1v_2}$, $m_{v_2v_3}$, and $m_{v_1v_3}$, or some combination of those.
For plotting convenience, in what follows we shall actually use the {\em squares} of those variables and denote them as
\beq
m_{12}^2 \equiv m^2_{v_1v_2}, \qquad m_{23}^2 \equiv m^2_{v_2v_3}, \qquad
m_{13}^2 \equiv m^2_{v_1v_3}. 
\label{mijdef}
\eeq
The variables (\ref{mijdef}) are in principle good candidates for the analysis, not only because they are Lorentz invariant, but also because their
distributions exhibit interesting kinematic features (edges and endpoints) which are traditionally used for determining the masses of the new particles 
$X_1$, $X_2$, $X_3$ and $\chi$ \cite{Hinchliffe:1996iu,Bachacou:1999zb,Hinchliffe:1999zc,Allanach:2000kt,Gjelsten:2004ki,Gjelsten:2005aw,Lester:2005je,Miller:2005zp,Burns:2009zi,Matchev:2009iw}. 

However, as discussed in refs.~\cite{Costanzo:2009mq,CL,Agrawal:2013uka,Kim:2015bnd,Debnath:2016gwz,Altunkaynak:2016bqe}, the multidimensional phase space $\left\{ m_{12}^2, m_{23}^2, m_{13}^2\right\}$
in this case in fact contains more information than is captured by edge-and-endpoint variables alone. As we will be describing in more detail in section~\ref{sec:theory}, 
the vicinity of the endpoints corresponds only to a fraction of the full boundary of the kinematically available phase space. 
This boundary is defined via the condition\footnote{Alternative parametrizations of the kinematic boundary can be found in \cite{Costanzo:2009mq,CL,Kim:2015bnd}.} $\DF=0$ where the variable $\DF$ will be introduced and defined in section~\ref{sec:theory} below. 
For now we simply remark that the location of this boundary contains the complete information about the spectrum in the cascade decay \cite{Costanzo:2009mq,CL}. 
A determination of this boundary (using Voronoi tessellations~\cite{Debnath:2015wra,Debnath:2016mwb}) has already been shown to result 
in an improvement in the measurement of the new physics mass spectrum~\cite{Debnath:2016gwz}.\footnote{For a related qualitative discussion, see page 573 in \cite{CL}.} More importantly, 
the phase space volume element has an enhancement near the boundary, even in the case of a compressed spectrum
\cite{Agrawal:2013uka}. 
This suggests that $\DF$ may be an effective discovery variable, especially in difficult scenarios of compressed spectra. 
The main goal of this paper will be to investigate the suitability of the $\DF$ variable as an analysis variable, either on its own, 
or when paired with the edge-and-endpoint variables\footnote{Note that $\DF$ is only defined for the phase space of four or more final state particles, and therefore cannot be used for the topologies in Fig.~\ref{fig:cascade}(a)-(c).}.

\begin{figure}
\begin{center}
\includegraphics[width=10cm]{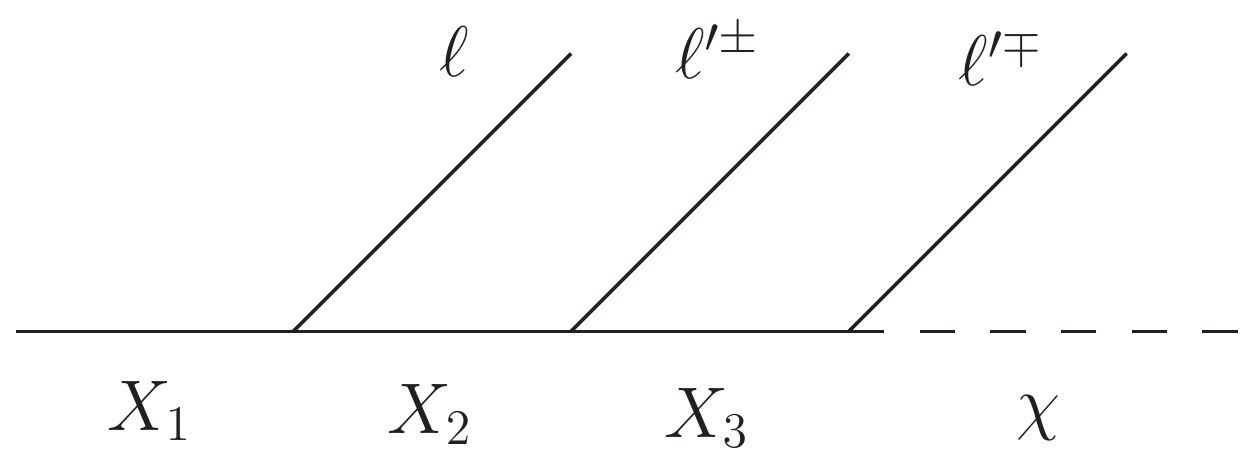}
\end{center}
\caption{The specific realization of the event topology from Fig.~\ref{fig:cascade}(d) which will be studied in this paper. Here $\ell'^{\pm}$ and $\ell'^{\mp}$ is a pair of opposite-sign, same-flavor leptons, while $\ell$ is a third lepton of a different flavor.}
\label{fig:lep}
\end{figure}

In order to demonstrate the basic idea, we adopt a specific realization of our benchmark decay topology from Fig.~\ref{fig:cascade}(d), by
specifying a final state on which we will base our analysis (see Fig.~\ref{fig:lep}). In particular, we will take $X_{1}$ and $X_{3}$ to be charged particles, while $X_{2}$ and $\chi$ are neutral. We also take the neutral particles to be flavor singlets. The SM particles produced in the second and third stages of the cascade are therefore oppositely charged, and have the same flavor, whereas the charge and flavor assignments of the SM particle produced in the first stage of the cascade are uncorrelated with the other two. Furthermore, in order to concentrate on what can be achieved using phase space techniques for discovery, we will aim to minimize possible complications due to challenging collider objects, so we choose the visible particles to be leptons. It is worth reiterating that our choice of final state is simply a choice of convenience in order to demonstrate the applicability of our methods, but the methods can be applied to photons, jets or even unstable SM particles with fully visible decays (such as visibly decaying $Z$-bosons) as well, at the potential cost of worse detector energy resolution and combinatorics. Our analysis {\it will} take into account the effect of finite energy resolution for leptons, as well as the combinatoric ambiguity about which lepton is emitted at the various decay stages. In particular, there will not in general be a way to distinguish which of the same-flavor, opposite-charge leptons is emitted higher upstream in the cascade. On the other hand, the lepton emitted in the first stage of the cascade {\it can} be distinguished by demanding it to carry 
a flavor different from
the same-flavor, opposite-charge lepton pair.

Since we aim to focus on improving signal selection in the case of compressed spectra, we adopt the following benchmark spectrum: $m_{X_{1}}=390$~GeV, $m_{X_{2}}=360$~GeV, $m_{X_{3}}=330$~GeV and $m_{\chi}=300$~GeV. Note that the choice of spectrum is mainly intended to demonstrate how well the kinematic variables in question compare to one another. Our conclusions would not be affected by raising all masses in the spectrum (while preserving the mass gaps), if we wanted to assign additional significance to this mass benchmark and avoid existing exclusion constraints for various potential underlying models, such as supersymmetry.

The outline of this paper is as follows: In the next section we will review the theoretical aspects of multidimensional phase space and formally introduce the $\DF$ variable. In section~\ref{sec:uniform}, we will then perform a preliminary study with simplified
assumptions to outline the salient features of $\DF$ as a discovery variable. In section~\ref{sec:massscan} we will address a subtlety about the use of a hypothesis spectrum in order to calculate $\DF$. Once this is done, we will then perform a realistic study of the performance of $\DF$ as a discovery variable in section~\ref{sec:SMBG}. We conclude in section~\ref{sec:conclusions}.


\section{Mathematical description of four-body phase space}
\label{sec:theory}

Let us start by introducing a {\it manifestly} Lorentz-invariant parametrization of the phase space for the cascade decay of our benchmark decay topology. Using the formalism of ref.~\cite{ByersYang},\footnote{For an alternative derivation, the curious reader is invited to follow Exercise 11 on page 574 in \cite{CL}.} we introduce the matrix 
\bea
{\mathcal Z}=\left\{ z_{ij}\right\}\quad {\rm with}\quad z_{ij}=
p_{i}\cdot p_{j}\, ,
\eea
where the $\{p_i\}$ are the four momenta of the final state particles $\ell$, $\ell'^{\pm}$, and $\chi$. The variables $\Delta_{i}$ can then be defined as
\bea
\det \left[\lambda I_{4\times4}-\mathcal{Z}\right]\equiv 
\lambda^4-\left(\sum_{i=1}^{4} \Delta_{i}\lambda^{4-i}\right).
\label{eq:definedeltas}
\eea
Among these variables, $\Delta_{4}$ will play a special role in the rest of this paper. As described in ref.~\cite{ByersYang}, the kinematically allowed region is given by $\Delta_{1,2,3,4}>0$, with the boundary located at\footnote{Alternative equivalent parametrizations of this kinematic boundary were previously derived in \cite{Costanzo:2009mq,CL,Kim:2015bnd}. However, those results were not used to study the {\em interior} of the kinematically allowed phase space, as we will be doing here.}
\bea
\Delta_4 =0,\qquad \Delta_{1,2,3}>0\,. \label{eq:boundary}
\eea
With the requirement that all $m_{ij}^{2}\ge 0$, outside of the kinematically allowed region the values of $\DF$ are negative and become arbitrarily large in magnitude as one moves towards infinity.

The general four-body phase space volume element is given by
\begin{equation} 
d\Pi_{4}=\left(\prod_{i<j}dm_{ij}^{2}\right)\frac{8}{(4\pi)^{10}M_{X_{1}}^{2}
\Delta_{4}^{1/2}}\ \delta\left(\sum_{i<j}m_{ij}^{2}-\left(M_{X_{1}}^{2} + 2 
\sum_{i=1}^{4} m_{i}^{2}\right)\right),
\label{4bps}
\end{equation}
where $m_{ij}^{2}=\left(p_{i}+p_{j}\right)^2=2z_{ij}+m_{i}^{2}+m_{j}^{2}$~\footnote{This is the general formula. For our analysis, while $m_{\chi}>0$, we will take the leptons to be massless.}. Note the factor of $\Delta_4^{-1/2}$, which causes an enhancement near the boundary $\DF=0$. 

Of course, the physically observable quantities depend not only on $d\Pi_{4}$ but on 
$|{\mathcal M}|^2$, the quantum mechanical matrix element squared for the decay:
\bea
d\Gamma=|{\mathcal M}|^{2}\ d\Pi_{4}\,.
\eea
In particular, for the benchmark decay topology of Fig.~\ref{fig:lep}, the volume element will be combined with the squares of the internal propagators in the cascade, which in the narrow width approximation are given as delta functions with arguments linear in the $m_{ij}^{2}$ and can therefore be used to perform some of the $m_{ij}^{2}$ integrals. As a result, the events fill out a three-dimensional phase space that can conveniently be fully parameterized in terms of the observables $m_{12}^{2}$, $m_{13}^{2}$ and $m_{23}^{2}$.

The enhancement in the phase space volume element near the boundary should make it clear why it is promising to consider $\DF$ as a discovery variable. The prominent features in the edge-and-endpoint variable distributions happen at the extremes of linear slicings of the three dimensional phase space, and therefore only a small fraction of signal events contribute to these features. In contrast, the prominent feature in the $\DF$ distribution at $\DF=0$ captures the full boundary of phase space, where the density of signal events is enhanced, so it is reasonable to expect that selecting for events near $\DF=0$, one could significantly enhance signal over background. 

It is worth remarking that the phase space for any known SM background process does not develop a singular structure like the one described in eq.~(\ref{4bps}). Furthermore, 
there is no reason to expect the $|{\mathcal M}|^2$ factor for the background to have any sharp features over the kinematically accessible signal region (the location of which depends on the signal spectrum). In particular, for a compressed signal spectrum which results in a relatively small signal region, the variation of the background matrix element over this region will in all likelihood be mild.

Note that for a given event, $\DF$ cannot be calculated from the observable data alone. As can be seen from eq.~\eqref{eq:definedeltas}, $\DF$ is equal to $-{\rm det}[{\mathcal Z}]$, and the last column and row of ${\mathcal Z}$ contain the four momentum of the lightest supersymmetric particle (LSP) $\chi$, which is unobservable. However, if one starts with a hypothesis for the spectrum $\{m_{X_{1}},m_{X_{2}},m_{X_{3}},m_{\chi}\}$, the on-shell constraints allow one to solve for all entries of ${\mathcal Z}$, and thus a mass hypothesis dependent value of $\DF$ can be calculated. The obvious question to ask then is whether this requirement for a spectrum hypothesis significantly weakens the usefulness of the $\DF$ variable. We will take up this question in section~\ref{sec:massscan}, drawing
the conclusion that $\DF$ is a powerful variable despite this caveat.


\section{Preliminary study with uniform background}
\label{sec:uniform}

In order to illustrate the usefulness of $\DF$, we wish to compare its performance as a discovery variable to the conventional edge-and-endpoint variables using the benchmark cascade decay and spectrum specified in the introduction. The performance of all variables will depend on the differential distribution of signal and background events, which as mentioned in the previous section will in turn depend on both the geometry of phase space as well as the matrix elements for signal and background. Again as emphasized in the previous section, the usefulness of $\DF$ originates from the phase space geometry for signal, in particular, the enhancement of the signal event density near the boundary of the kinematically allowed region where there is no strong reason to expect a feature in the density of background events. Therefore, we devote this section to a toy study where we minimize the effects of the matrix elements and of the background event distribution, by taking all particles in the signal decay chain to be scalars, and we make the highly simplifying approximation that the background varies not only slowly over the signal region but is in fact uniformly distributed over phase space (parameterized in terms of the coordinates $m_{ij}^{2}$). We will also use the true signal spectrum in calculating $\DF$ and return to the issue of having to scan over spectrum hypotheses in the next section, before we do a full analysis with SM backgrounds and a signal model 
with spins of new particles assigned SUSY-like in section~\ref{sec:SMBG}.

Since we use a uniformly distributed background, we need to define a finite box 
in the three-dimensional space formed by
the 
three
$m_{ij}^{2}$ variables in order to deal with only a finite number of background events. We choose the box size as twice the maximal possible signal value in each of the $m_{ij}^{2}$ variables. This choice ensures that finite energy resolution in the detector does not push signal events outside the box, and that no artificial features are introduced in background distributions at small but negative values of $\DF$, close to but outside the signal region. We generate high statistics samples with one million signal and background events each, where in the signal the flavors of the leptons $\ell$ and $\ell'$ are randomly assigned as electrons or muons. We only consider events where those two flavors are distinct.

Even in this preliminary study, we will need to face two complications. One is finite energy resolution, as mentioned, while the other complication arises from combinatoric ambiguities. Note that in our benchmark topology of Fig.~\ref{fig:lep}, it cannot be experimentally determined in which order the particles $\ell^{'+}$ and $\ell^{'-}$ are emitted in the cascade, leading to a combinatoric ambiguity. As argued in ref.~\cite{Allanach:2000kt}, in such a case it is advantageous to work with {\it ordered} variables instead, so we define and work with the variables
\beq
m_{1(hi)}^{2}\equiv {\rm max}(m_{12}^{2},m_{13}^{2}),\qquad m_{1(lo)}^{2}\equiv{\rm min}(m_{12}^{2},m_{13}^{2}). \label{eq:orderedm}
\eeq
Note that there is no 
combinatorial ambiguity in defining $m_{23}^{2}$ as we require $\ell'$ and $\ell$ to have distinct flavors. Due to the combinatorial ambiguity, there are two possible values of $\DF$ for every event, and both of them will be used when populating $\DF$ histograms.
In setting up our study, we will choose to start by using perfect energy resolution and by ignoring the combinatoric ambiguity, before introducing them below. We do this because there are a few important lessons we can learn even before the analysis is made more complicated by these effects.

\begin{figure}
\begin{center}
\includegraphics[width=0.75 \textwidth]{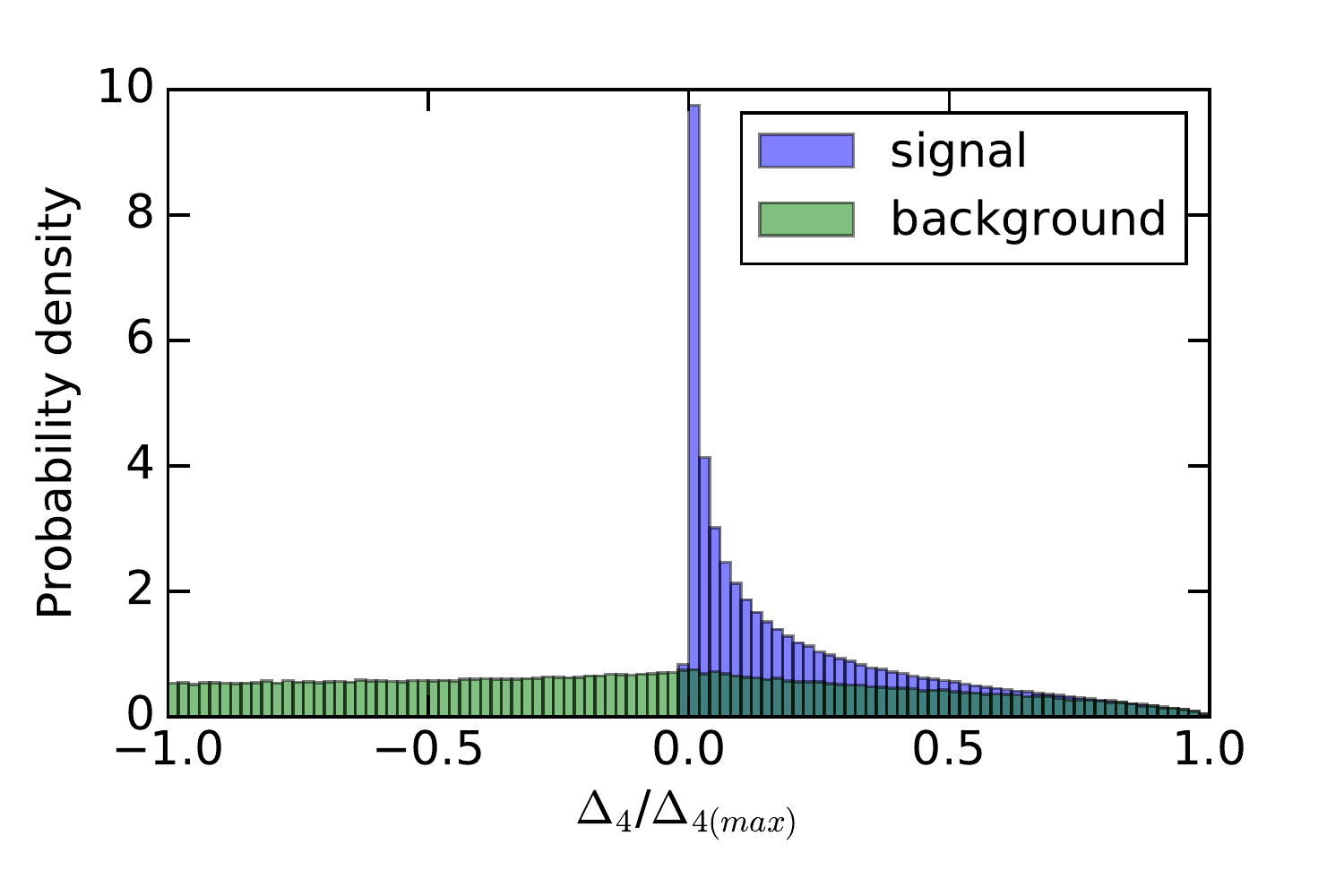}
\end{center}
\caption{The $\DF$ histograms for signal 
(blue)
and (uniformly distributed) background
(green). The distributions are normalized by the maximum $\DF$ value for the chosen mass spectrum, $(m_{X_1}, m_{X_2}, m_{X_3}, m_\chi)=(390, 360, 330, 300)$ GeV.
The feature in the background distribution near $\DF=0$ is caused by the volume between constant $\DF$ surfaces becoming maximal.
}
\label{fig:delta4hist}
\end{figure}

As mentioned in the introduction, an ideal discovery variable that eliminates the need for precise background modeling would exhibit a strong feature in the distribution of the signal while the background distribution is smooth at the same position, such that a sideband analysis can pick out the signal as in a bump-hunting analysis. At first sight, $\DF$ seems to be a promising variable along these lines, since the signal event density is enhanced near $\DF=0$ while the background event density has no reason to be enhanced at the same surface, the location of which after all is dependent on the signal spectrum. Unfortunately, this line of thinking misses a potential problem, namely that even though the {\it density} of background events may be smooth near the surface $\DF=0$, the phase space in which signal and background events are distributed is three-dimensional, and in making a one-dimensional histogram of $\DF$, one has to integrate the phase space volume between surfaces of constant $\DF$. This can still introduce a feature into the background $\DF$ histogram if the volume between contours itself exhibits a feature near $\DF=0$. This does in fact happen to be the case, since the gradient of $\DF$ is small on a significant portion of the boundary surface, increasing the volume between $\DF$ contours there. The resulting $\DF$ histogram for signal and background (uniform density) is shown in Fig.~\ref{fig:delta4hist}, where the normalization of the signal and background histograms has been chosen such that they both contain the same total number of events. 
Here $\DF$ values are normalized by the maximum $\DF$ for the chosen mass spectrum, $(m_{X_1}, m_{X_2}, m_{X_3}, m_\chi)=(390, 360, 330, 300)$ GeV.
When the number of background events are significantly higher than the number of signal events, as is often the case for searches for new physics, and when the distributions become smeared due to finite energy resolution, the presence of the background feature at $\DF=0$ will make a simple bump hunt based on a sideband analysis difficult, since the signal can be misinterpreted as a background systematic \cite{CL}.

We therefore switch to a different approach for a search strategy. In order to compare the effectiveness of the different variables in selecting signal events, we construct a performance curve of each variable as follows\footnote{The spirit of these curves is similar to a receiving operator characteristic (ROC) curve, even though they are not technically ROC curves.}. For a given variable, a histogram is made of the signal and background. For the $m^{2}$ variables, the interval of interest in the histogram is between the maximum and minimum possible values predicted by the spectrum, and for $\DF$ it is the interval between $\pm\Delta_{4(max)}$, also as predicted by the spectrum. The interval of interest is divided into 100 bins\footnote{We verify that the procedure outlined here is not sensitive to the choice of binning.}. The first entry in the performance curve is the ratio of signal to background events (S/B) in the bin with the highest number of signal events. To obtain the second entry in the performance curve, this bin is combined with the bin to its left or to its right, whichever of the two has the larger number of signal events, and S/B is calculated for the combined two-bin region. For the third entry in the performance curve, these two bins are combined with the neighboring bin with the higher number of signal events, and so on. The procedure stops when all bins containing signal events are exhausted, and therefore the last entry in the performance curve corresponds to S/B over the full signal region for the variable in question. Note that the ordering of the bins in terms of signal events (as opposed to S/B) reduces the reliance on background modeling.

\begin{figure}
\begin{center}
\includegraphics[width=0.99 \textwidth]{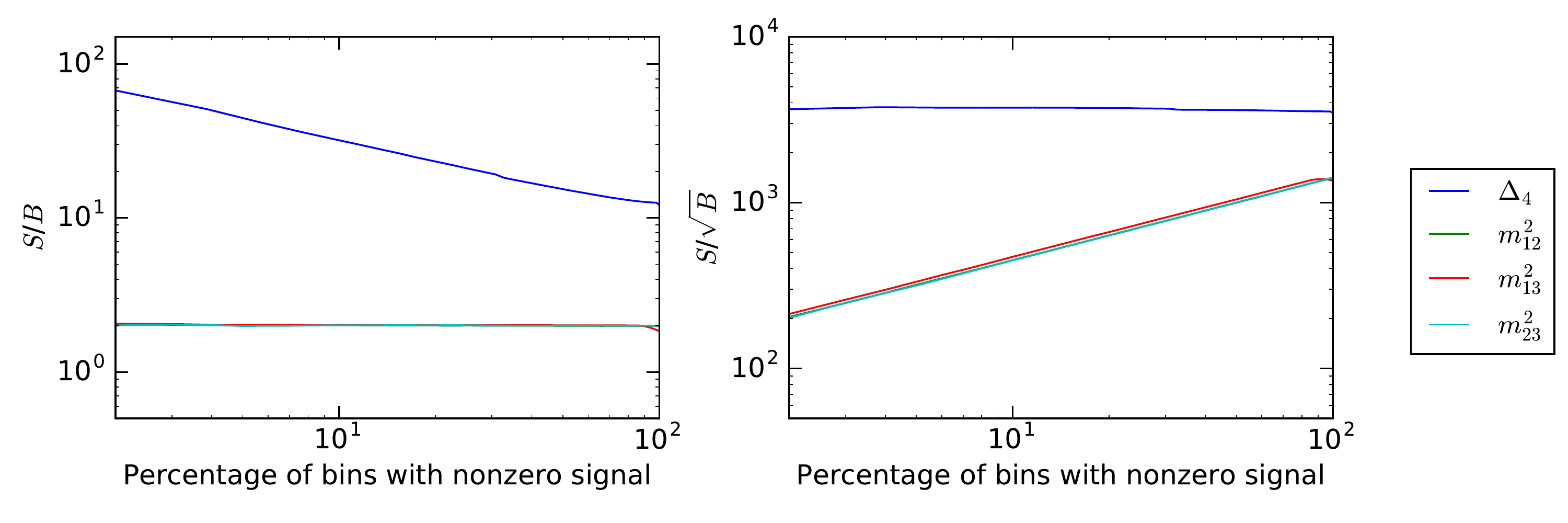}
\end{center}
\caption{Performance curves for $\DF$ and the invariant mass variables using the ${\rm S}/{\rm B}$ (left panel) and ${\rm S}/\sqrt{{\rm B}}$ (right panel) metrics, with perfect energy resolution. See 
the main text for the way in which we construct these curves.
}
\label{fig:ROCuniformsinglenosmear}
\end{figure}

We point out that the performance curves of any two variables may be meaningfully compared independently of the overall signal and background normalizations, since any change in the signal and background normalizations will multiply the performance curve of all variables by the same common factor. Using the same procedure, for completeness we also produce performance curves for the ${\rm S}/\sqrt{{\rm B}}$ metric\footnote{S/B and ${\rm S}/\sqrt{{\rm B}}$ are the relevant quantities measuring signal significance in searches that are systematics and statistics dominated, respectively, and we wish to remain agnostic as to which case may apply in the experimental search of interest. }. These performance curves are shown in Fig.~\ref{fig:ROCuniformsinglenosmear}. Note that by construction, the background has a flat distribution in all $m_{ij}^{2}$ variables, and in the absence of spin correlations, the signal has an exactly flat distribution in $m_{12}^{2}$ and $m_{23}^{2}$, and a nearly flat distribution in $m_{13}^{2}$ as well. This explains the near-flatness of the S/B performance curves of the $m_{ij}^{2}$ variables, as well as the $\sqrt{N_{\rm bins}}$ scaling for the ${\rm S}/\sqrt{{\rm B}}$ performance curves. As can be seen from the figures, $\DF$ performs significantly better than these with respect to both metrics.

\begin{figure}
\begin{center}
\includegraphics[width=\textwidth]{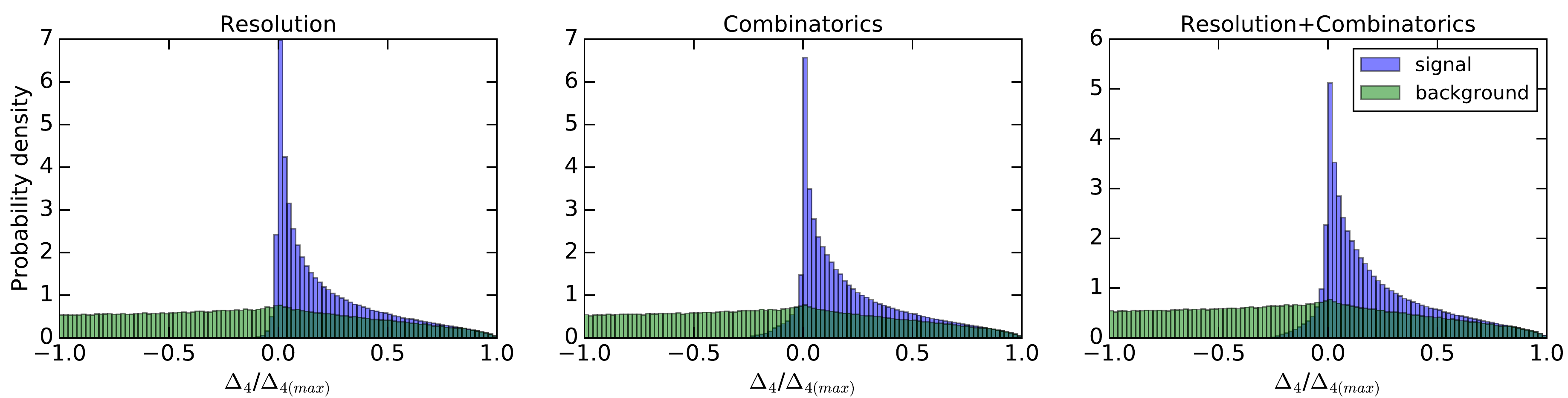}
\end{center}
\caption{The $\DF$ histograms for signal 
(blue)
and (uniformly distributed) background
(green), with energy resolution and combinatoric ambiguities included. To be compared to Fig.~\ref{fig:delta4hist}}
\label{fig:smearedhistogram}
\end{figure}

Encouraged by this result, we proceed to check whether it is robust in the presence of finite detector energy resolution and combinatorial ambiguities. We use the EM calorimeter resolution based on the CMS-TDR~\cite{Bayatian:2006nff} 
\beq
\frac{\sigma_E}{E}=(0.0026)\oplus \frac{0.0363~{\rm GeV}^{1/2}}{\sqrt{E}}\oplus\frac{0.124~{\rm GeV}}{E}, \label{eq:CMSres}
\eeq
where the energy $E$ is defined in GeV. For the muon resolution we utilized values (in terms of muon momentum and pseudorapidity) summarized in Figure 1.5 of the CMS-TDR~\cite{Bayatian:2006nff}.
Since the background 
that we consider in this preliminary study
is not physical and has no four-vectors associated with it, we leave it unmodified. To incorporate combinatorial ambiguities into the analysis, we use the ordered $m^{2}$ variables as defined 
in eq.~\eqref{eq:orderedm}, and we populate $\DF$ histograms by both possible values for each event as mentioned above. The effect of smearing and combinatorics on the $\DF$ distribution of figure~\ref{fig:delta4hist} is shown in Fig.~\ref{fig:smearedhistogram}.

\begin{figure}
\begin{center}
\includegraphics[width=\textwidth]{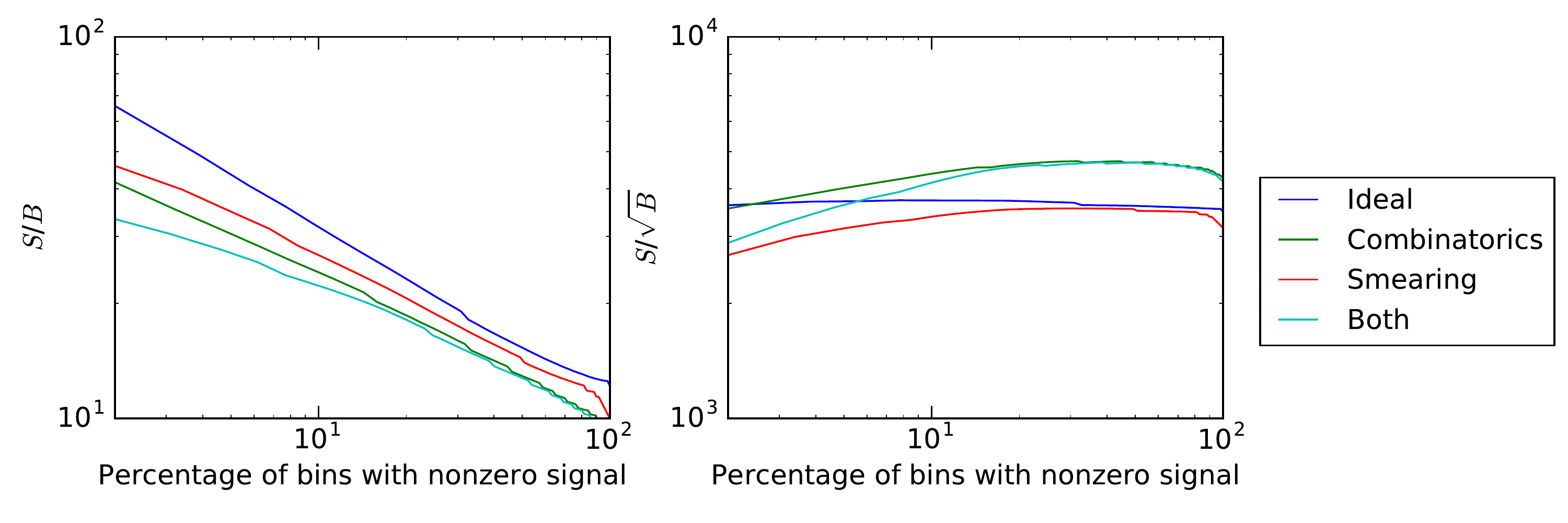}
\end{center}
\caption{The effect of energy resolution and combinatorics on the significance performance curve of $\DF$ is shown using the ${\rm S}/{\rm B}$ (left panel) and ${\rm S}/\sqrt{{\rm B}}$ (right panel) metrics.}
\label{fig:ROCdelta4degradation}
\end{figure}

\begin{figure}
\begin{center}
\includegraphics[width=0.99 \textwidth]{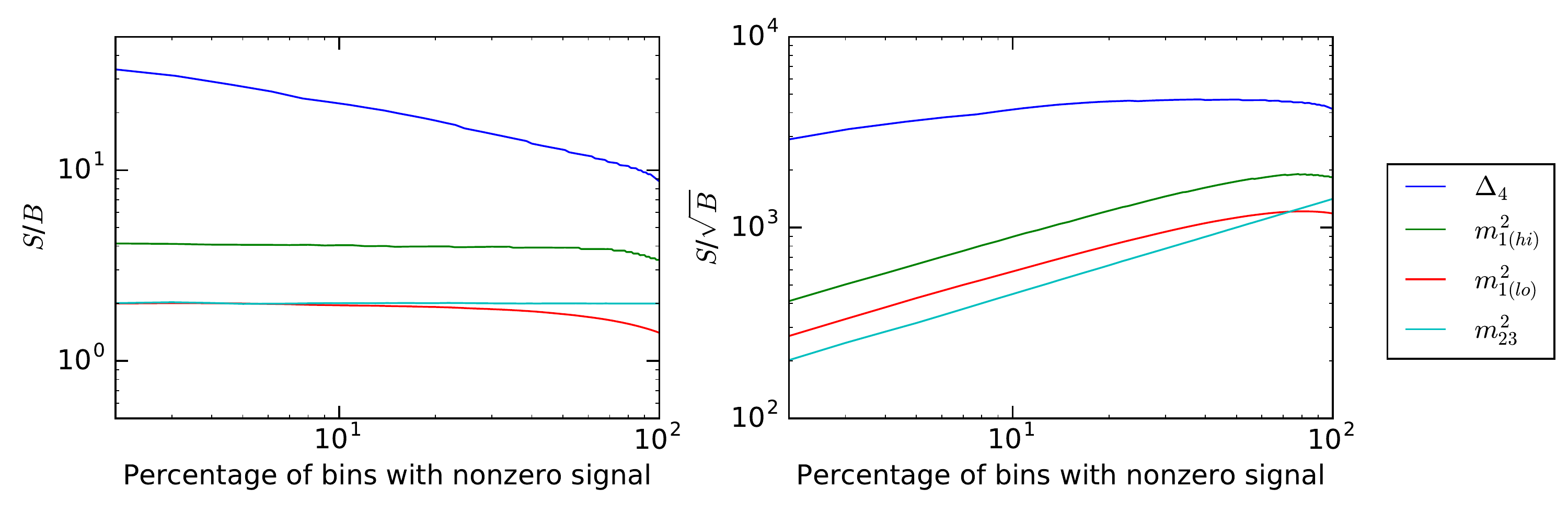}
\end{center}
\caption{
The same as Fig.~\ref{fig:ROCuniformsinglenosmear}, but
taking the finite energy resolution and combinatoric effects into account.
}
\label{fig:ROC_uniform_single_smeared}
\end{figure}

As a result of both smearing and combinatorics, the performance curves for $\DF$ in Fig.~\ref{fig:ROCuniformsinglenosmear} are 
mildly
degraded, which can be seen in Fig.~\ref{fig:ROCdelta4degradation}. In Fig.~\ref{fig:ROC_uniform_single_smeared}, the performance curves of $\DF$ and the edge-and-endpoint variables are compared with energy resolution and combinatorics included. $\DF$ is seen to still outperform the edge-and-endpoint variables, but by a smaller margin.

\begin{figure}
\begin{center}
\includegraphics[width=0.99 \textwidth]{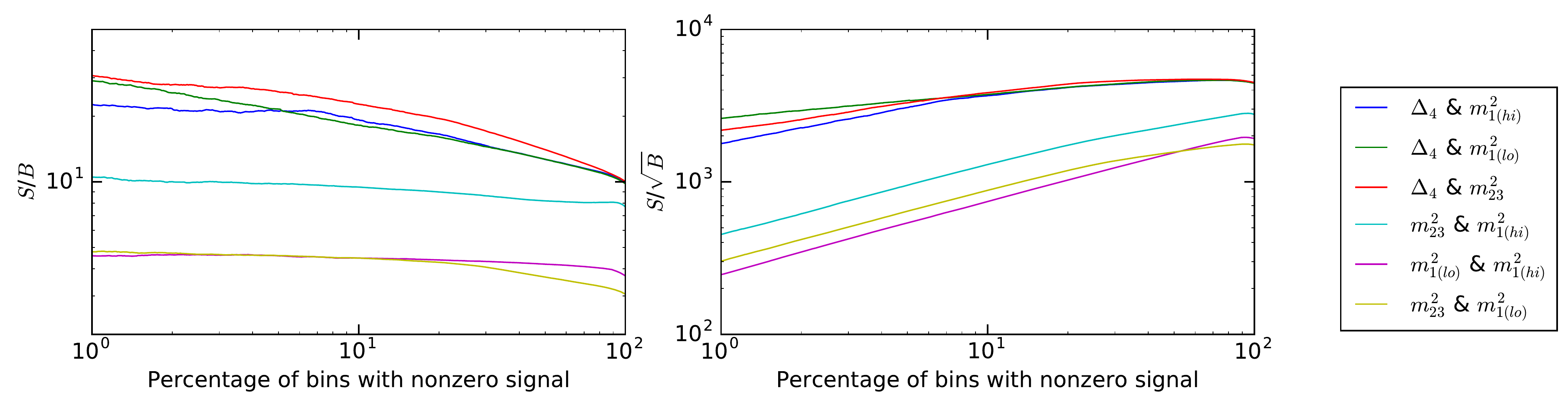}
\end{center}
\caption{Performance curves for {\em pairs of} variables among $\DF$ and the invariant mass variables, using the ${\rm S}/{\rm B}$ 
(left panel)
and ${\rm S}/\sqrt{{\rm B}}$ (right panel)
metrics, taking finite energy resolution and combinatoric effects into account.
}
\label{fig:ROC_uniform_pair_smeared}
\end{figure}

After this preliminary comparison among single kinematic variables as discovery tools, it is also interesting to look at how well {\it pairs} of variables compare to one another.  In particular we will be interested in whether pairing $\DF$ with the $m^{2}$ variables will be more effective than pairing one of the $m^{2}$ variables with another one. The procedure we use to perform this comparison closely mirrors the procedure outlined above for the case of a single variable. In particular, for any pair of variables, signal and background events populate a double histogram in the two variables in question (the same binning parameters are used in each variable as described earlier in this section). The (double) bins are then ordered in order of their signal contribution, but {\it without} demanding that the bins that are combined neighbor one another, and performance curves of S/B and of S/$\sqrt{{\rm B}}$ are made. The effects of both smearing and of combinatorics are included. 
We exhibit the results in Fig.~\ref{fig:ROC_uniform_pair_smeared} from which
it is easy to see that variable pairs including $\DF$ perform better than variable pairs not including $\DF$ with respect to both metrics.

\section{Scanning over mass spectra}
\label{sec:massscan}

Encouraged by the promising results of our preliminary study described in the previous section, we will devote this section to address the issue of the spectrum dependence in calculating $\DF$. In particular, since the true signal spectrum is not known a priori, analyses involving $\DF$ will need to scan over all possible (correctly ordered) spectra. Below, we will show that the significance is maximized at least locally for the true spectrum, a result which is consistent with the conclusions of ref.~\cite{Debnath:2016gwz}. Therefore, if one were to scan over all spectra and use the spectrum that yields the highest significance, then the performance curve based on the true spectrum offers a guaranteed, and in fact potentially conservative (should other spectra exist far from the true spectrum that lead to even higher significance), benchmark for comparison against the performance curves of the $m^{2}$ variables. The significances we report will be {\it local}. The calculation of a global significance requires the use of a trials factor which is tricky to define for this analysis and is beyond the scope of this paper.


The question of the potential existence of other local (or even global) maxima of significance requires extensive calculational resources, since a fine scan over four masses is required\footnote{We expect such resources to be available to the LHC collaborations, however most the analysis in this paper is performed entirely on standalone computers.}. However, since we will show below that the true spectrum yields at least a local maximum, with a high significance value, then if other local maxima with even higher significance should exist, this would only strengthen the discovery potential, not reduce it, but at the cost of having to give up the claim that the spectrum can be simultaneously measured in the same analysis. We will therefore not make this claim in this study.

To demonstrate that the true spectrum yields a local maximum of significance, we will compare the performance curves of $\DF$ for a range of hypothesized spectra obtained by {\it local} deformations around the true spectrum.  A background uniform in the $m_{ij}^{2}$ variables is used as in the previous section, and finite energy resolution as well as combinatorial ambiguities are included in the analysis.

\begin{figure}
\begin{center}
\includegraphics[width=0.99 \textwidth]{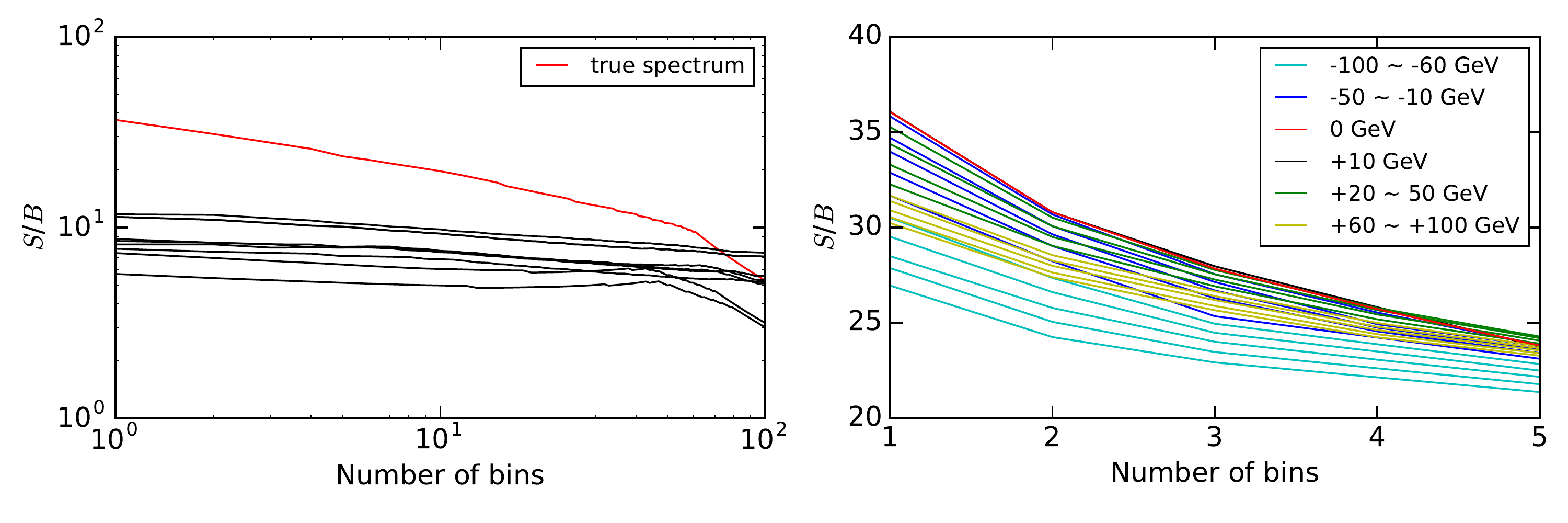}
\end{center}
\caption{Performance curves for $\DF$ calculated by using a range of hypothesis spectra and the S/B metric. 
Left: Each one of the plotted curves  corresponds to deforming the spectrum by changing each of the four masses up or down by $\pm$ 10 GeV. For comparison, 
the red curve highlights the true spectrum.
Right: Each one of the curves corresponds to deforming the spectrum along the flat direction described in the main text over a wide range. The color scheme corresponds to the change in the mass of the LSP.
}
\label{fig:spectrum_scan}
\end{figure}

For the local scan near the true spectrum, we allow each of the four masses to change up or down by 10 GeV, resulting in 8 variations. The performance curves obtained as a result of the scan are shown on the left-hand side of Fig.~\ref{fig:spectrum_scan}. It is easy to see that for any low or moderate number of bins in the performance curve, the true spectrum yields the highest significance. The strong reduction in the performance as one goes away from the true spectrum (along any direction other than the flat direction, see the next paragraph) can be traced to the fact that the sharp peak at $\DF=0$ is only present when $\DF$ is calculated for the true spectrum, and is severely distorted otherwise, thereby erasing the most distinctive feature in the signal distribution compared to the background distribution.

We also perform a finer one-dimensional scan along a special direction. In particular, while the $m_{ij}^{2}$ variables are sensitive to changes in the mass gaps in the spectrum, there is a direction where the endpoints of all three $m_{ij}^{2}$ distributions remain fixed. We parameterize this direction in terms of the change in the mass of the LSP from its benchmark value. As shown in ref.~\cite{Debnath:2016gwz}, $\DF$ is sensitive to changes along the flat direction, while the effect on the shape of the $m_{ij}^{2}$ distributions is minimal. These results are shown in the right-hand side of Fig.~\ref{fig:spectrum_scan}, with the conclusion that small deformations along the flat direction leave the performance curve unchanged (within statistical errors) while more substantial deformations reduce the significance. The results of the scans presented above thus confirm our claim that the $\DF$ performance has a local maximum for the true spectrum.


\section{Study with SM background}
\label{sec:SMBG}

Having obtained encouraging results in our toy study with uniform background, and having dealt with the subtlety of scanning over spectrum hypotheses in calculating $\DF$, we 
are now in the position to conduct
a much more realistic study, with SM backgrounds, matrix element effects in the signal, finite detector resolution, and combinatorics
taken into consideration.
For the signal, we consider a benchmark model where $X_{1}$ is a scalar muon partner, $X_{2}$ is a heavy fermion, $X_{3}$ is a scalar electron partner, and $\chi$ is the fermionic LSP. 
It should be emphasized again
that we are not arguing for this as a signal model to be taken literally; as argued in the introduction, this model is chosen to make an apples-to-apples comparison between $\DF$ and the $m^{2}$ variables possible, without introducing distracting complications. 
Nevertheless, we believe that our proposed analysis is straightforwardly applicable to the SUSY signal searches in the channel we study here. This signal model guarantees the flavor arrangement of the three leptons in our benchmark cascade. The dominant SM background for this final state is $WZ^{(*)}$ production 
followed by their leptonic decays.
Since our benchmark spectrum ensures that the opposite sign, same flavor lepton pair invariant mass remains well below $m_{Z}$, we impose a $Z$-veto in simulating the background, so that the region with off-shell $Z$'s can be scanned efficiently.

We perform our parton-level simulation for signal and background using \texttt{MG5@aMC}~\cite{Alwall:2014hca}, and apply energy resolution for final state leptons according to the CMS-TDR~\cite{Bayatian:2006nff} [see also eq.~\eqref{eq:CMSres}]. 
We use the following selection cuts on the events:
\beq
p_{T,\ell} > 10 \textrm{ GeV},\,\,\,\,\, \left| \eta_{\ell} \right| <2.5,\,\,\,\,\, \Delta R_{\ell\ell}\ge0.4, \,\,\,\,\, 15 \textrm{GeV}< m_{\ell^+\ell^-} < 65 \textrm{ GeV} \,\,\,\,\,(\ell = e,\, \mu).
\eeq
Here the invariant mass cut in the second line is relevant only to same-flavor opposite-sign lepton pairs. 

\begin{figure}
\begin{center}
\includegraphics[width=\textwidth]{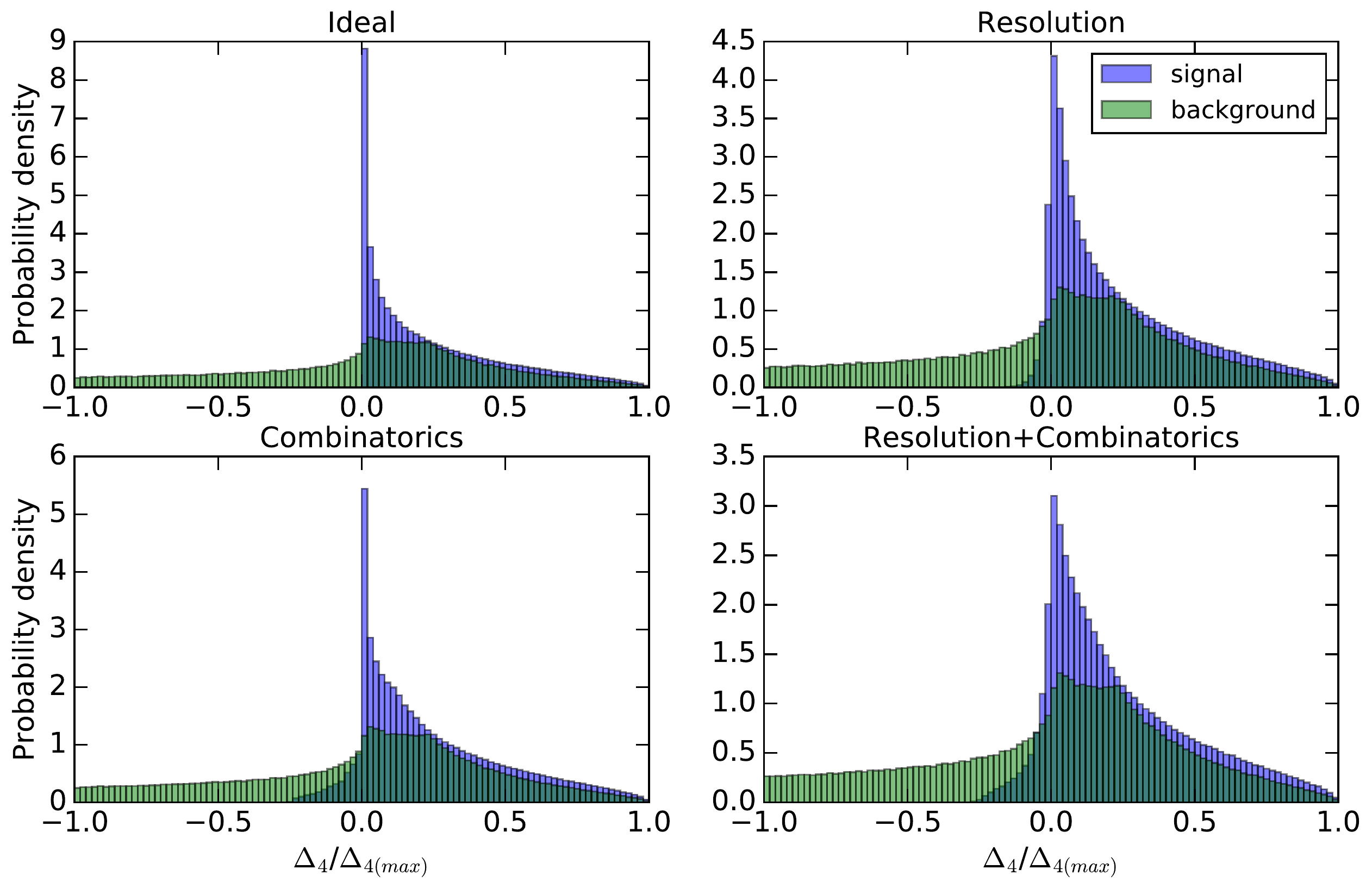}
\end{center}
\caption{The $\DF$ histograms for signal 
(blue)
and the SM background
(green), with energy resolution and combinatoric ambiguities included.}
\label{fig:smearedhistogramSM}
\end{figure}

\begin{figure}
\begin{center}
\includegraphics[width=\textwidth]{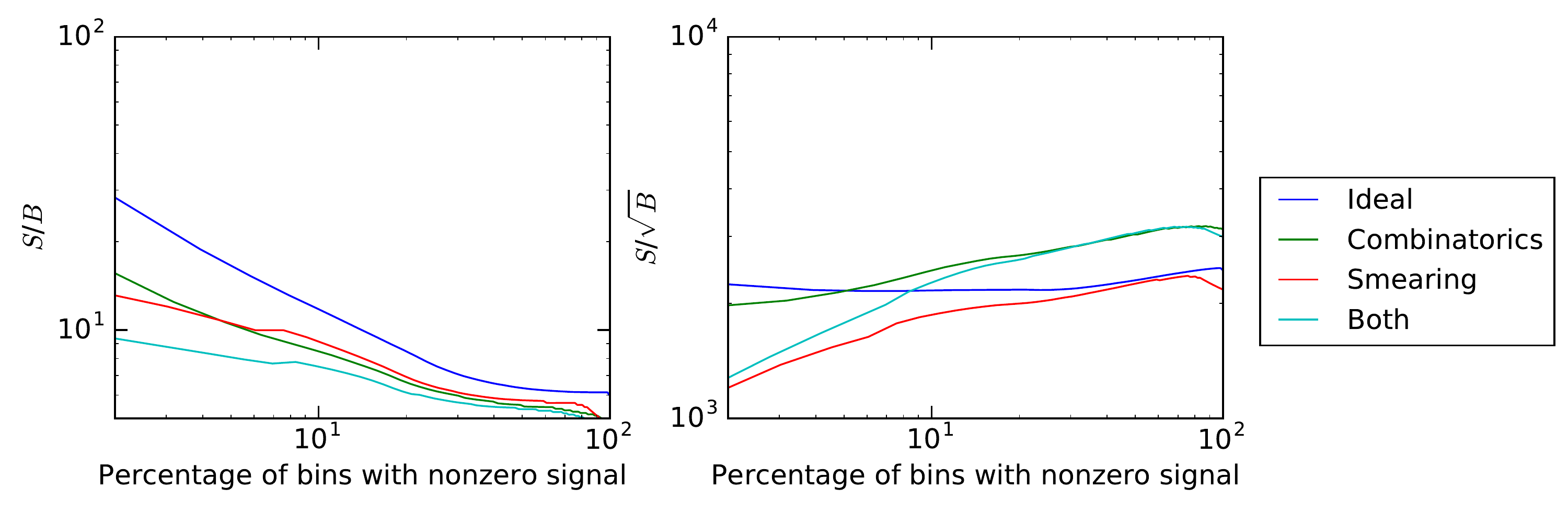}
\end{center}
\caption{The effect of energy resolution and combinatorics on the significance performance curve of $\DF$ is shown using the ${\rm S}/{\rm B}$ (left panel) and ${\rm S}/\sqrt{{\rm B}}$ (right panel) metrics.}
\label{fig:ROCdelta4degradationSM}
\end{figure}

\begin{figure}
\begin{center}
\includegraphics[width=0.99 \textwidth]{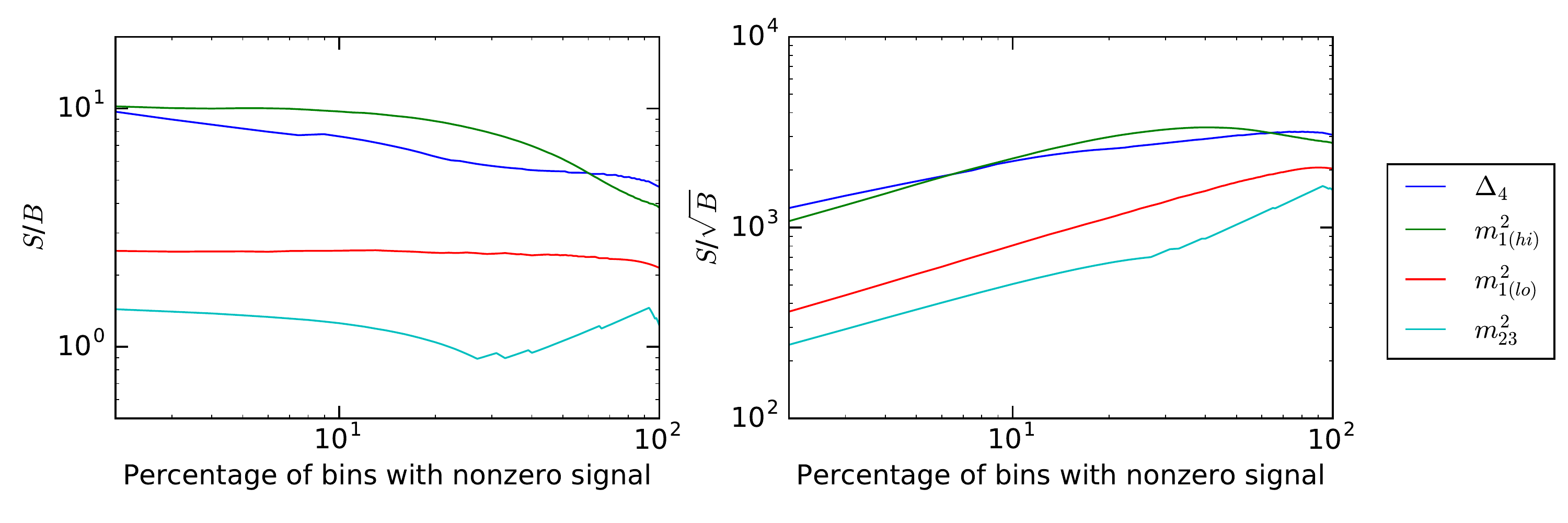}
\end{center}
\caption{Performance curves for $\DF$ and the $m^{2}$ variables using the ${\rm S}/{\rm B}$ and ${\rm S}/\sqrt{{\rm B}}$ metrics.
}
\label{fig:ROC_SM_single_smeared}
\end{figure}

For the generated signal and background event samples, we plot the $\DF$ distributions, as well as the effect of smearing and combinatorics on these distributions, in Fig.~\ref{fig:smearedhistogramSM}. The resulting performance curves for $\DF$ are obtained following the same steps as in section~\ref{sec:uniform}, and shown in Fig.~\ref{fig:ROCdelta4degradationSM}. We then compare the performance of $\DF$ to the edge-and-endpoint variables in Fig.~\ref{fig:ROC_SM_single_smeared}.
We observe that the $\DF$ variable becomes less powerful
than it was in our preliminary study with uniform background. 
The main reason for this 
degradation is because
the matrix elements and the parton distribution functions that govern the phase space distribution of SM background events lead more events to lie close to the regions in which $\DF$ is 
smaller than that for
the uniform background distribution \cite{CL};
for example, the event population in the same-flavor lepton pair invariant mass is enhanced at small values due to the mixing between $\gamma$ and $Z$, resulting in more background population at small values of $\DF$.
Nonetheless, $\DF$ shows a comparable performance to the strongest $m^{2}$ variable with respect to both metrics.

\begin{figure}
\begin{center}
\includegraphics[width=0.99 \textwidth]{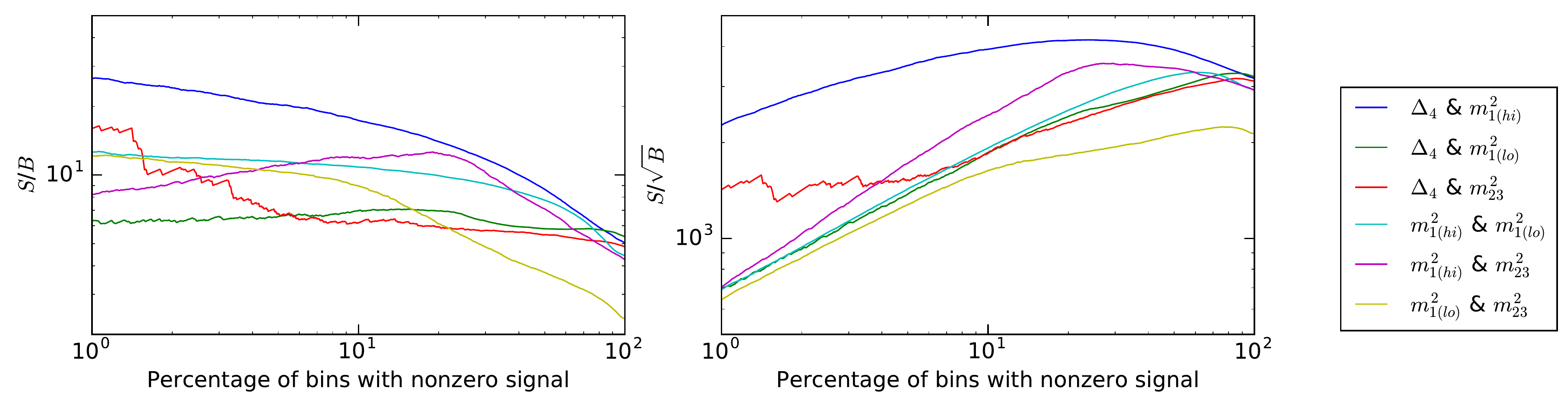}
\end{center}
\caption{Performance curves for pairs of variables among $\DF$ and the $m^{2}$ variables, using the ${\rm S}/{\rm B}$ and ${\rm S}/\sqrt{{\rm B}}$ metrics.
}
\label{fig:ROC_SM_pair_smeared}
\end{figure}

Furthermore, as we pointed out in our preliminary exercise, some $m^2$ variable, when combined with $\DF$, may outperform traditional approaches with $m^2$ variables only.
Indeed, the same expectation goes through for the signal under consideration, which is supported by the results presented in Fig.~\ref{fig:ROC_SM_pair_smeared}. 
As one would expect based on the single variable results of Fig.~\ref{fig:ROC_SM_single_smeared}, the best performance is achieved by the combination between $m_{1(hi)}^2$ and $\DF$ (blue lines) in both the S/B (left panel) and the S/$\sqrt{\rm B}$ (right panel) metrics.
Therefore, we find that $\DF$ can play, at least, a {\it complementary} role in separating signal from background, hence expediting a discovery of new physics.

\section{Conclusions}
\label{sec:conclusions}

As we approach the end of Run II in the LHC experiment, the absence of a discovery of new physics makes it increasingly more imperative to focus on scenarios where a new physics signal may exist in the data, but not be distinctive enough to register in searches looking for high momentum particles. This happens for example when the new particles that are produced decay in a cascade with a compressed spectrum. We argued that using the variable $\DF$, which arises naturally in describing four-body phase space, allows one to design a search strategy in such a scenario that is quite inclusive and does not rely strongly on background modeling.\footnote{In place of $\Delta_4$, one could in principle also use the geometrical distance to the kinematical boundary (\ref{eq:boundary}), a possibility which was entertained in \cite{CL}. However, that choice has disadvantages: the geometrical distance is suboptimal in terms of performance and cannot be easily computed by analytical means.} We do this by focusing our attention on only the part of the event containing the cascade decay, using Lorentz-invariant variables, and by not using detailed properties of the background in designing our search strategy. We have argued that even though the calculation of $\DF$ requires a hypothesis for the mass spectrum in the cascade decay, the significance has a local maximum for the true signal spectrum which can be used as a benchmark of comparison against the performance of other variables. We have compared the performance of the variable $\DF$, both singly and paired with conventional edge-and-endpoint variables, in a study using SM backgrounds, spin correlations, finite energy resolution and combinatoric effects, concluding that $\DF$ can significantly enhance the signal both for systematics-dominated (S/B metric) and statistics-dominated (S/$\sqrt{\rm B}$ metric) searches.

\acknowledgments
The research of the authors is supported by the 
National Science Foundation Grant Number PHY-1620610 and
by the Department of Energy under Grants DE-SC0010296 and DE-SC0010504. 
DK was supported in part by the Korean Research Foundation (KRF) through the CERN-Korea Fellowship program, and is presently supported by the Department of Energy under Grant No. DE-FG02-13ER41976/DE-SC0009913.

\end{document}